\documentclass[11pt,twoside]{article}

\usepackage{epsfig,amsfonts,color}
\usepackage{amsmath}

\usepackage{amssymb, palatino, geometry,url}
\usepackage{algorithmic}
\usepackage[noresetcount,lined,boxed]{algorithm2e} 

\usepackage[colorlinks=true,linkcolor=blue,citecolor=blue,urlcolor=blue]{hyperref}

\usepackage{fancyhdr}
\newcommand{\be}{\begin{equation}}
\newcommand{\ee}{\end{equation}}
\newcommand{\bea}{\begin{eqnarray}}
\newcommand{\eea}{\end{eqnarray}}

\newcommand{\bvec}{\left(\begin{array}{c}}
\newcommand{\evec}{\end{array}\right)}
\newcommand{\bsub}{\begin{subequations}}
\newcommand{\esub}{\end{subequations}}

\usepackage{lineno}

\usepackage{amsthm}

\usepackage{appendix}
\usepackage{multirow}

\usepackage{caption}
\usepackage{subcaption}

\begin{document}

\title{Quantifying Space-Time Load Shifting\\  Flexibility in Electricity Markets}

\author{Weiqi Zhang${}^{\ddag}$ and Victor M. Zavala${}^{\ddag}$\thanks{Corresponding Author: victor.zavala@wisc.edu}\\
  {\small ${}^{\ddag}$Department of Chemical and Biological Engineering}\\
 {\small \;University of Wisconsin-Madison, 1415 Engineering Dr, Madison, WI 53706, USA}}
 \date{}
\maketitle

\vspace{-0.2in}

\begin{abstract}
The power grid is undergoing significant restructuring driven by the adoption of wind/solar power and the incorporation of new flexible technologies that can shift load in space and time (e.g., data centers, battery storage, and modular manufacturing). Load shifting is needed to mitigate space-time fluctuations associated with wind/solar power and other disruptions (e.g., extreme weather). The impact of load shifting on electricity markets is typically quantified via sensitivity analysis, which aims to assess impacts in terms of price volatility and total welfare. This sensitivity approach does not explicitly quantify operational flexibility (e.g., range or probability of feasible operation). In this work, we present a computational framework to enable this; specifically, we quantify operational flexibility by assessing how much uncertainty in net loads (which capture uncertain power injections/withdrawals) can be tolerated by the system under varying levels of load shifting capacity. The proposed framework combines optimization formulations that quantify operational flexibility with power grid models that capture load shifting in the form of virtual links (pathways that transfer load across space-time). Our case studies reveal that adding a single virtual link that shifts load in either space or time can lead to dramatic improvements in system-wide flexibility; this is because shifting relieves space-time congestion that results from transmission constraints and generator ramping constraints. Our results provide insights into how the incorporation of flexible technologies can lead to non-intuitive, system-wide gains in flexibility. 
\end{abstract}


\section{Introduction}

The power grid is undergoing major structural changes as incorporation of renewable energy introduces a high level of spatio-temporal variability and uncertainty. Under this trend, leveraging flexibility from various technologies is key to achieving effective and efficient decarbonization for the power grid \cite{babatunde2020power}. Specifically, load shifting has been widely studied as a mechanism that can help mitigate variability and uncertainty  \cite{alstone20172025}. Examples of shiftable loads include electric vehicles \cite{rotering2010optimal,schuller2015quantifying}, batteries \cite{sioshansi2009estimating}, and buildings \cite{hao2012demand}.  Flexibility is normally defined based on the ability to shift loads temporally, but this notion can also be extended to capture geographical flexibility, which can be enabled by spatially-distributed systems such as data centers \cite{liu2011greening} and modular manufacturing facilities (e.g., ammonia) \cite{shao2021mitigating}. 

While harnessing shifting flexibility is necessary, current energy market designs are not fully capable of remunerating shifting flexibility directly. For instance, traditional electricity markets for demand response focus on remunerating peak shaving or load shedding \cite{albadi2007demand, jordehi2019optimisation}. Recently, a pricing scheme for shiftable loads has been proposed to provide a direct incentive for load-shifting flexibility, which is computed without considering power network models \cite{werner2021pricing}. An electricity market design has also been recently proposed to directly price and remunerate load-shifting flexibility \cite{zhang2020flexibility,zhang2022remunerating}. Here, shifting flexibility is captured in the form of virtual links, which are non-physical pathways that transfer power across space and time. The virtual link concept has also been used to remunerate flexibility provided by energy storage systems (e.g., load/power shifting is represented as a temporal virtual shift) \cite{zhang2022pricing}. Impacts of flexibility on power grid performance are typically studied via sensitivity analysis and use a wide variety of metrics  \cite{heggarty2020quantifying}. Unfortunately, sensitivity approaches can be computationally intensive and do not directly quantify operational flexibility (i.e., ability of the system to maintain feasible operation).  

Flexibility analysis has been widely studied in the field of process systems engineering, dating back from the seminal work by Grossmann and co-workers \cite{halemane1983optimal}. In this framework, flexibility is defined as the ability for a system to maintain feasibility (satisfy all its constraints) given a set of uncertain parameters/data. This approach has been applied to various systems such as chemical processes \cite{pistikopoulos1995novel}, power systems \cite{cui2021network, ulbig2015analyzing}, and autonomous vehicles \cite{wang2009autonomous}.   
Operational flexibilty can be quantified using deterministic and stochastic approaches. The deterministic approach proposed in \cite{grossmann1983optimization} measures flexibility by taking a robust optimization view of the problem (i.e. identifying worst-case uncertainty that can be tolerated) \cite{zhang2016relation}. Pulsipher and co-workers have recently shown that this approach can be extended to capture different types of uncertainty representations  (e.g., ellipsoidal) \cite{pulsipher2019computational}. The stochastic approach defines flexibility in terms of probability that the system retains feasible operation and this requires sampling/quadrature approaches to compute probabilities \cite{straub1990integrated}. A benefit of the deterministic approach is that it bypasses the need for sampling and can capture a large number of uncertain parameters; however, this approach is restricted to convex constraints. On the other hand, the stochastic approach can accommodate nonconvex constraints but might require large number of samples to enable accurate computations.  

In this work, we apply the deterministic flexibility analysis framework of Pulsipher and co-workers to an electricity market clearing formulation that captures load shifting flexibility in the form of virtual links \cite{zhang2020flexibility}. We explore the question of measuring/quantifying the improvement in flexibility as a result of adding virtual links systems. Our results show that adding a {\em single pair} of virtual links can dramatically expand the feasible space of the power grid system in non-intuitive ways, resulting in large gains in flexibility (up to 250\% for a spatial shifting case and 80\% for a temporal shifting case). We also explore the cost of flexibility provision by considering economic/budget constraints; the results show that modest increases in operating cost can result in substantial improvements in flexibility. Overall, we believe that our framework can provide valuable insights into how the deployment of new flexible technologies can help increase flexibility and mitigate increasing levels of uncertainty that power grids are facing. The framework can also be used to identify what types of technologies (and how many) might be needed to achieve various flexibility levels.  

The manuscript is structured as follows. Section \ref{sec:market} goes over the market clearing framework that we study. Section \ref{sec:flex_analysis} reviews the flexibility analysis framework applied. Section \ref{sec:case_study} presents numerical results of the case studies. Section \ref{sec:conclusion} concludes the paper.

\section{Power Grid Model with Virtual Links}
\label{sec:market}
\begin{figure}[!htp]
    \centering
    \includegraphics[width=0.75\textwidth]{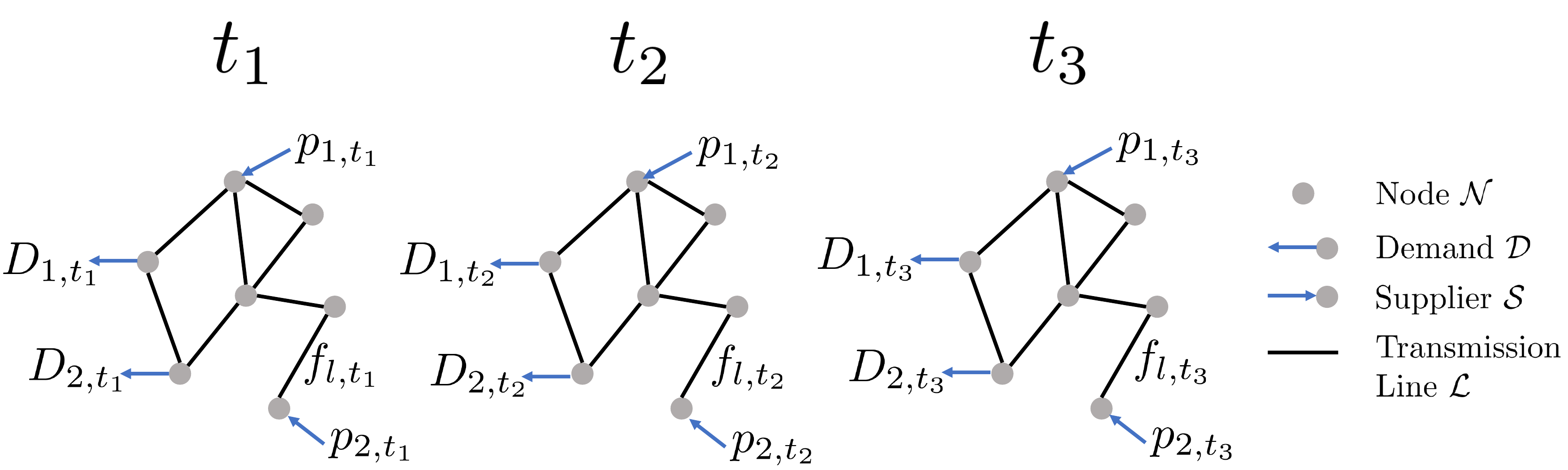}
    \caption{\small Illustration of a base electricity market system with three time intervals. }
    \label{fig:market}
\end{figure}

In this paper, we consider the economic dispatch formulation presented in \cite{zhang2020flexibility, zhang2022remunerating}. This formulation can be interpreted as a market clearing formulation. Similar formulations (known as coordinated management) are also applied to environmental supply chain studies \cite{sampat2019coordinated}. Independent system operators (ISOs) solve these types of formulations to determine optimal power allocations to stakeholders. The market system setup is visualized in Figure \ref{fig:market}; this considers a set of suppliers (owner of generators) $\mathcal{S}$ and consumers (owners of loads) $\mathcal{D}$ connected to a transmission network comprised of geographical nodes $\mathcal{N}$ and transmission lines $\mathcal{L}$ (owned by transmission service providers).
\\

Each supplier $i \in \mathcal{S}$ is connected to the power grid at node $n(i)\in \mathcal{N}$. The supplier offers available capacity $\bar{p}_{i,t} \in [0, \infty)$ and ramping capacity $\Delta p_i \in [0, \bar{p}_i]$. The set of suppliers at node $n$ is defined as $\mathcal{S}_n := \{i \in \mathcal{S} \, | \, n(i) = n\} \subseteq \mathcal{S}$. The market will decide the cleared allocation $p_{i,t}$ (amount of power injected) for each supplier $i\in \mathcal{S}$ at time $t\in\mathcal{T}$. We use $p$ to denote the collection of all cleared allocations. 
\\

The transmission network consists of the node set $\mathcal{N}$ and a set of transmission lines $\mathcal{L}$. Each line $l \in \mathcal{L}$ is associated with a sending node $\mathrm{snd}(l) \in \mathcal{N}$ and receiving node $\mathrm{rec}(l) \in \mathcal{N}$. The definitions of $\mathrm{snd}(l)$ and $\mathrm{rec}(l)$ are interchangeable because power can flow in either direction. For each node $n \in \mathcal{N}$, we define its set of receiving lines $\mathcal{L}_n^{\textrm{rec}} := \{l \in \mathcal{L} \, | \, n=\textrm{rec}(l)\} \subseteq \mathcal{L}$ and its set of sending lines $\mathcal{L}_n^{\textrm{snd}} := \{l \in \mathcal{L} \, | \, n=\textrm{snd}(l)\}\subseteq \mathcal{L}$. The problems identifies flows $f_{l,t}$ that satisfy capacity bounds $f_{l,t} \in [-\bar{f}_l, \bar{f}_l]$ and direct-current (DC) power flow equations:
\begin{equation}
\label{eq:dc_flow}
f_{l,t} = B_l(\theta_{\textrm{snd}(l),t} - \theta_{\textrm{rec}(l),t}),
\end{equation}
where $B_l \in \mathbb{R}_+$ is the line susceptance and $\theta_{n,t} \in \mathbb{R}$ is the phase angle at node $n \in \mathcal{N}$. We note that the DC power flow model is a linear model that approximates the actual power flow physics under small phase angle difference conditions. We use this linear model to avoid quadratic constraints in formulation \eqref{opt:F_mip}, which make the flexibility analysis much more computationally expensive.
\\

We model consumers as a net aggregated demand $D_{n,t} \in [0,+\infty)$ at node $n$ and time $t$. The net demand considers injection of renewable power, and are thus treated as uncertain parameters. In this paper, we assume that loads cannot be curtailed and thus the amount of net demand $D_{n,t}$ must be satisfied. However, this does not mean that the loads are inflexible; specifically, we will see that the loads can be shifted to alternative space-time locations and be served at such locations. 
\\

The base electricity market clearing framework is as follows:
\begin{subequations}
\label{opt:market_base_model}
\begin{align}
&\underset{p,f,\theta,\delta}{\text{min}} & & \sum_{t \in \mathcal{T}} \sum_{i \in \mathcal{S}} \alpha_{i,t}^p p_{i,t} \label{opt:market_base_model_obj} \\
& {\text{s.t.}} & & \sum_{l \in \mathcal{L}_n^{\textrm{rec}}} f_{l,t} + \sum_{i \in \mathcal{S}_n} p_{i,t} = \sum_{l \in \mathcal{L}_n^{\textrm{snd}}} f_{l,t} + D_{n,t}, \quad  n \in \mathcal{N}, t \in \mathcal{T} \label{opt:market_base_model_balance} \\ 
&&& f_{l,t} = B_l(\theta_{\textrm{snd}(l),t} - \theta_{\textrm{rec}(l),t}), \quad l \in \mathcal{L}, t \in \mathcal{T} \label{opt:market_base_model_dc} \\
&&& -\Delta \bar{p}_i \leq p_{i,t} - p_{i,t-1} \leq \Delta \bar{p}_i, \quad i \in \mathcal{G}, t \in [2,3,...,T] \label{opt:market_base_model_ramping}\\
&&& 0 \leq p \leq \bar{p},\, -\bar{f} \leq f \leq \bar{f} \label{opt:market_base_model_capacity} 
\end{align}
\end{subequations}
Here, the objective function \eqref{opt:market_base_model_obj} is the total generation cost. 
Constraints \eqref{opt:market_base_model_balance} capture power balance at each space time node. Constraints \eqref{opt:market_base_model_dc} capture the DC power flow model. 
Constraints \eqref{opt:market_base_model_ramping} capture ramping constraints for each generator.
Constraints \eqref{opt:market_base_model_capacity} capture the capacity for generators and transmission lines.
Note that the ramping capacities of generators determine how much temporal flexibility the system has; if generators have small ramping capacities, the system will not able to respond to time-varying demands.  In addition, note that transmission capacities determine how much spatial flexibility the system has, but the effect is more complex as this also involved the network topology/connectivity. The combination of the physical model, ramping capacities, and transmission capacities define the operational flexibility of the system (i.e., its ability to absorb demands). 
\\

\begin{figure}[!htp]
    \centering
    \includegraphics[width=0.75\textwidth]{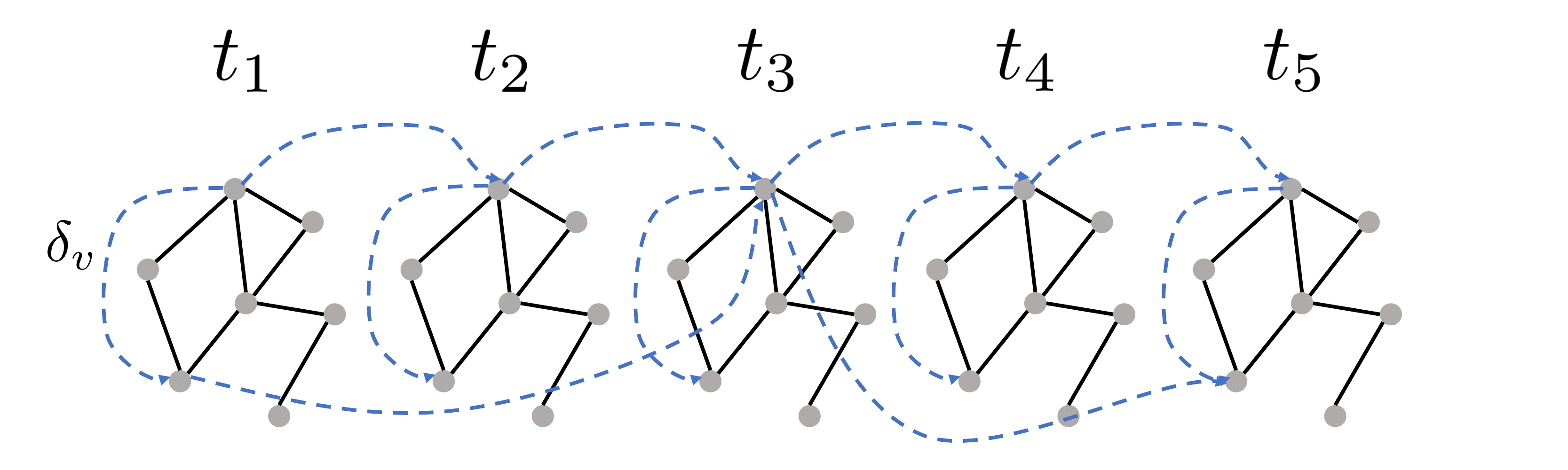}
    \caption{\small Illustration of an electricity market system with a set of virtual links $\mathcal{V}$, with five time intervals. The set of virtual links form a space-time network that acts as an additional infrastructure layer to match electricity demand and supply, on top of the transmission network. }
    \label{fig:network}
\end{figure}

Formulation \eqref{opt:market_base_model} does not capture shifting flexibility that emerging technologies (i.e. data centers, energy storage, and distributed manufacturing facilities) can offer. Zhang et al. \cite{zhang2020flexibility} extended this formulation to capture flexibility in the form of virtual links. Virtual links are a modeling abstraction that captures space-time shifting flexibility from different types of technologies. In general, virtual links represent the non-physical transfer of power injection/extraction from space-time location $(n,t)$ to another $(n',t')$. Note that over one virtual link, it is allowed to have either $t = t'$ or $n = n'$ to capture purely spatial or temporal shifting, but the equalities cannot hold simultaneously. The set of virtual links can be seen as an additional infrastructure layer (on top of the physical system) that can be leveraged to respond to satisfy demands, as visualized in Figure \ref{fig:network}. 
\\

The formulation considers a set of virtual links $\mathcal{V}$; each virtual link $v \in \mathcal{V}$ is associated with a sending space-time node $\text{snd}(v) = (n_{\text{snd}(v)}, t_{\text{snd}(v)})$ and a receiving space-time node $\text{rec}(v) = (n_{\text{rec}(v)}, t_{\text{rec}(v)})$. We define $\mathcal{V}_{n,t}^\textrm{snd} := \{v\in\mathcal{V} \,|\, \textrm{snd}(v) = (n,t)\} \subseteq \mathcal{V}$, $\mathcal{V}_{n,t}^\textrm{rec} := \{v\in\mathcal{V} \,|\, \textrm{rec}(v) = (n,t)\} \subseteq \mathcal{V}$ to be the set of sending and receiving virtual links at space-time node $(n,t)$. This setting captures the special case in which $v$ is a spatial virtual link if it connects nodes at different locations but same time ($n_{\text{snd}(v)} \neq n_{\text{rec}(v)}, t_{\text{snd}(v)} = t_{\text{rec}(v)}$), and the special case in which $v$ is a temporal virtual link if it connects nodes at different times but at the same location ($t_{\text{snd}(v)} \neq t_{\text{rec}(v)}, n_{\text{snd}(v)} = n_{\text{rec}(v)}$). The clearing formulation needs to decide the virtual link allocation $\delta_v$ (amount of power injection/extraction shifted) subject to the operation model. Incorporating virtual links leads to the general clearing formulation shown below:
\begin{subequations}
\label{opt:market_model}
\begin{align}
&\underset{p,f,\theta,\delta}{\text{min}} & & \sum_{t \in \mathcal{T}} \sum_{i \in \mathcal{S}} \alpha_{i,t}^p p_{i,t} + \sum_{v \in \mathcal{V}} \alpha^\delta_{v} \delta_v \label{opt:market_model_obj} \\
& {\text{s.t.}} & & \sum_{l \in \mathcal{L}_n^{\textrm{rec}}} f_{l,t} + \sum_{i \in \mathcal{S}_n} p_{i,t} + \sum_{v \in \mathcal{V}^{\textrm{snd}}_{n,t}} \delta_v = \sum_{l \in \mathcal{L}_n^{\textrm{snd}}} f_{l,t} + \sum_{v \in \mathcal{V}^{\textrm{rec}}_{n,t}} \delta_v + D_{n,t}, \quad  n \in \mathcal{N}, t \in \mathcal{T} \label{opt:market_model_balance} \\ 
&&& f_{l,t} = B_l(\theta_{\textrm{snd}(l),t} - \theta_{\textrm{rec}(l),t}), \quad l \in \mathcal{L}, t \in \mathcal{T} \label{opt:market_model_dc} \\
&&& -\Delta \bar{p}_i \leq p_{i,t} - p_{i,t-1} \leq \Delta \bar{p}_i, \quad i \in \mathcal{G}, t \in [2,3,...,T] \label{opt:market_model_ramping}\\
&&& 0 \leq p \leq \bar{p},\, -\bar{f} \leq f \leq \bar{f} \label{opt:market_model_capacity} \\ 
&&& c_E(\delta) = 0; \, c_I(\delta) \geq 0 \label{opt:market_model_vl_constraints}
\end{align}
\end{subequations}
The objective function \eqref{opt:market_model_obj} is the total cost, consisting of generation costs and virtual link costs. 
Power balance constraints \eqref{opt:market_model_balance} now capture the effect of virtual links. In addition to constraints \eqref{opt:market_model_dc}-\eqref{opt:market_model_capacity}, constraints \eqref{opt:market_model_vl_constraints} capture a set of equality and inequality constraints that model the operations of virtual link providers. Specific forms of constraints depend on the characteristics of flexibility providers. For instance, the constraints for virtual links provided by a data center can be modeled as follows \cite{zhang2020flexibility}:
\begin{subequations}
\label{eq:dc_vl}
\begin{gather}
D_{n,t} + \sum_{v\in\mathcal{V}^{\textrm{rec}}_{n,t}}\delta_v - \sum_{v\in\mathcal{V}^{\textrm{snd}}_{n,t}} \delta_v \geq 0, \quad  n \in \mathcal{N}, t \in \mathcal{T} \label{eq:dc_vl_lb}\\
0 \leq \delta \leq \bar{\delta} \label{eq:dc_vl_capacity}
\end{gather}
\end{subequations}
Here, constraints \eqref{eq:dc_vl_lb} capture the bound on the realized load at each data center site. Constraints \eqref{eq:dc_vl_capacity}  capture the limit on the amount of loads allowed to shift via each virtual link. We note that, in general, the virtual link allocation variables $\delta_v$ are allowed to be either positive or negative. A positive value means that the virtual link is shifting power extraction (e.g., load shifting), while a negative value means that the virtual link is shifting power injection (e.g., can be done using batteries). 

\section{Flexibility Analysis Framework}
\label{sec:flex_analysis}
To quantify the flexibility in a given market system \eqref{opt:market_model}, we need a framework to compute a scalar value which measures the flexibility of a market system given a configuration (or lack thereof) of virtual links. This will allow us to compare how flexible the system is with different levels of load-shifting flexibility captured in the form of virtual links. In this work we apply the flexibility analysis framework proposed by Pulsipher et al. \cite{pulsipher2019computational} to derive an optimization  formulation for the system. The framework computes the so-called flexibility index (denoted as $F$); this index measures the size of uncertainty under which system operation is feasible (satisfies all constraints). Here, we briefly review key concepts of the framework. 

Consider a general system defined by the constraint set:
\begin{subequations}
\label{eq:system}
\begin{alignat}{2}
g_j(x;\xi,\bar{\delta}) &\leq 0, \quad j \in \mathcal{J}^I \\ 
h_j(x;\xi,\bar{\delta}) &= 0, \quad j \in \mathcal{J}^{E}
\end{alignat}
\end{subequations}
where $g_j(\cdot), j \in \mathcal{J}^I$ are inequality constraint functions and $h_j(\cdot), j \in \mathcal{J}^E$ equality constraint functions, $x$ the decision variables, $\xi$ uncertain parameters, and $\bar{\delta}$ system design parameters. Once the design parameters $\bar{\delta}$ are fixed, the feasibility of system \eqref{eq:system} under a given realization $\xi$ can be computed as
\begin{subequations}
\label{opt:feasibility}
\begin{alignat}{2}
& \psi(\xi) := \quad & \underset{x,t}{\min} \quad & t\\ 
&& {\text{s.t.}} \quad & g_j(x;\xi,\bar{\delta}) \leq t,\quad j \in \mathcal{J}^I \\
&&& h_j(x;\xi,\bar{\delta}) = 0, \quad j \in \mathcal{J}^{E}
\end{alignat}
\end{subequations}
For a given realization $\xi$, the system is feasible if $\psi(\xi) \leq 0$ (meaning there exists a solution $x$ that satisfies all constraints in \eqref{eq:system}), and infeasible otherwise. The feasible set of the system can then be defined as $\Xi := \{\xi \, | \, \psi(\xi) \leq 0\}$. 
\\

The flexibility index problem seeks to identify the largest uncertainty set $T(\alpha)$, parameterized by a scalar $\alpha \in \mathbb{R}_+$ such that system \eqref{eq:system} remains feasible for all possible realizations of the uncertain parameters $\xi$ in $T(\alpha)$. 
The size of the uncertainty set $T(\alpha)$ scales with $\alpha$. The uncertainty set needs to have a pre-defined shape and a nominal point (denoted as $\bar{\xi}$) so that it can be parameterized by a single scalar $\alpha$. Common uncertainty sets are shown in Table \ref{tab:sets}. 

\begin{table}[!htp]
\centering
\caption{Common representations of uncertainty sets.}\label{tab:sets}
\begin{tabular}{ |c|c| }
 \hline
 Shape & Uncertainty set \\ 
 \hline
 Hyperbox & $T_{box}(\alpha) = \{\xi \,|\,\bar{\xi} - \alpha \Delta \xi \leq \xi \leq \bar{\xi} - \alpha \Delta \xi \}$ \\ 
 Ellipsoid & $T_{box}(\alpha) = \{\xi \,|\, (\xi - \bar{\xi})^T V^{-1} (\xi - \bar{\xi}) \leq \alpha \}$ \\
 $\ell_p$ norm set & $T_{box}(\alpha) = \{\xi \,|\, \|\xi - \bar{\xi}\|_p \leq \alpha \}$ \\
 \hline
\end{tabular}
\end{table}

A natural choice of uncertainty set representation and nominal point is highly dependent on the specific applications considered. Normally, the hyperbox set is the default option for flexibility analysis as it is intuitive, requires minimal data, and is computationally inexpensive (it can be expressed as a set of linear constraints). However, the hyperbox set is not necessarily the best option for all applications; for instance, in cases where uncertain parameters are correlated in space-time (e.g., wind power or loads) it is more appropriate to use ellipsoidal sets.
\\

\begin{figure}[!htp]
    \centering
    \includegraphics[width=0.4\textwidth,trim={20 15 20 10},clip]{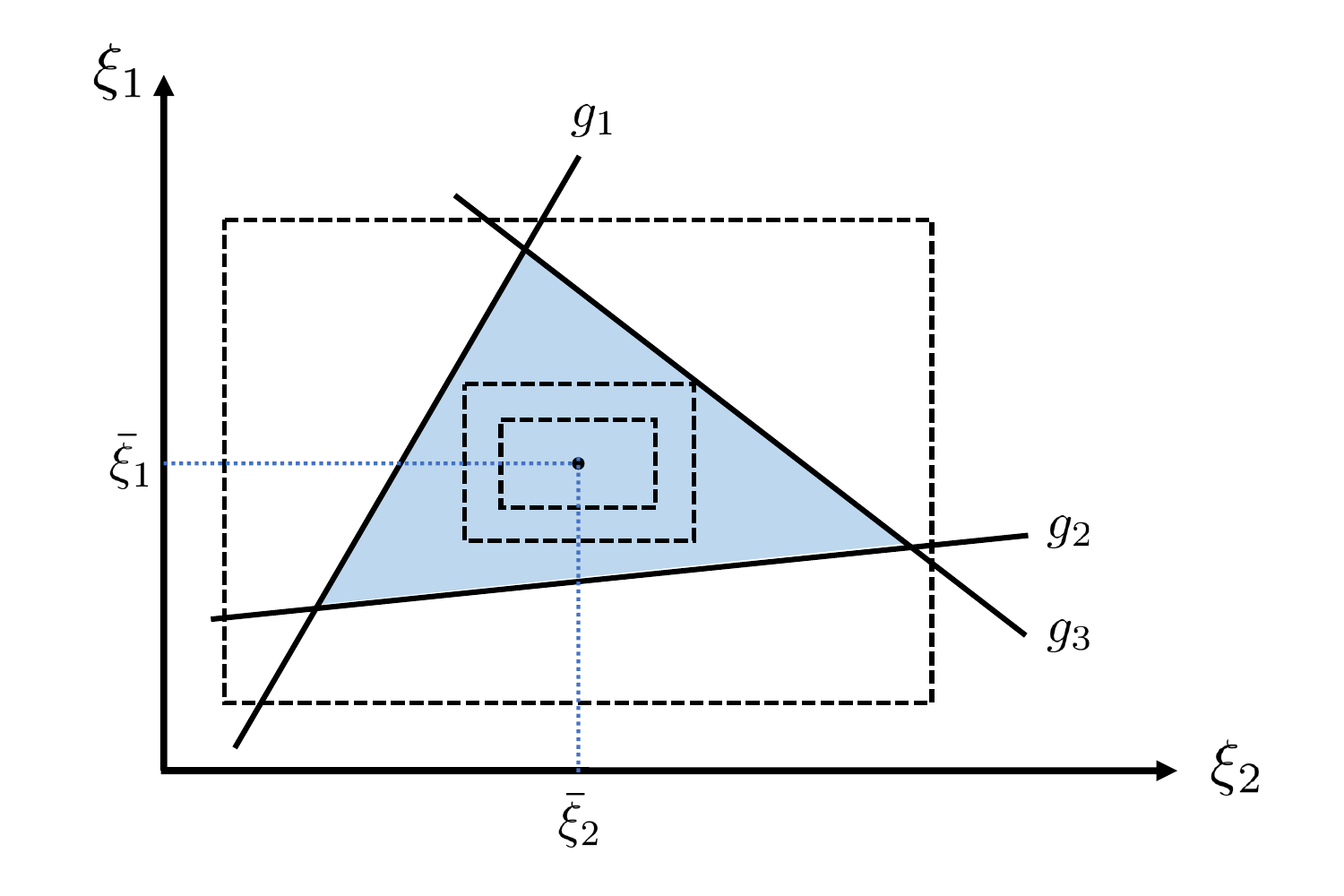}
    \caption{\small Illustration of hyperbox set for a system with linear constraints. Dashed boxes denote hyperbox sets of different $\alpha$ values centered at the same nominal point. Any feasible set that touches the constraints will be as large as the optimal feasible set, and any feasible set that can be fit in the system cannot be larger than the optimal feasible set. }
    \label{fig:feasible_set}
\end{figure}

The flexibility index $F$ is defined as : 
\begin{subequations}
\label{opt:F}
\begin{alignat}{2}
& F := \quad & \underset{\alpha\in\mathbb{R}_+}{\max} \quad & \alpha \\ 
&& {\text{s.t.}} \quad & \underset{\xi \in T(\alpha)}{\max} \psi(\xi) \leq 0
\end{alignat}
\end{subequations}
Problem \eqref{opt:F} seeks to find the largest uncertainty set $T(\alpha)$ such that the system \eqref{eq:system} attains at least one feasible solution for all realizations of $\xi$. Problem \eqref{opt:F} is a tri-level optimization problem and is generally challenging to solve. However, \cite{swaney1985index} has shown that problem \eqref{opt:F} is equivalent to finding the minimum $\alpha$ along the boundary of the feasible set $\Xi$ if $T(\alpha)$ is compact and constraint functions $g_j(x;\xi,\bar{\delta})$ and $h_j(x;\xi,\bar{\delta})$ are Lipschitz continuous in $x, \xi, \bar{\delta}$. Note that the boundary of the feasible set can be re-written as $\partial \Xi = \{\xi \, | \, \psi(\xi) = 0\}$. The flexibility index problem can thus be written as:
\begin{subequations}
\label{opt:F_alt}
\begin{alignat}{2}
& F = \quad & \underset{\alpha\in\mathbb{R}_+, \xi \in T(\alpha)}{\min} \quad & \alpha \\ 
&& {\text{s.t.}} \quad & \psi(\xi) = 0 \label{opt:F_alt_con}
\end{alignat}
\end{subequations}
Figure \ref{fig:feasible_set} illustrates a hyperbox feasible set for a system with linear constraints. This gives an intuitive explanation on why formulations \eqref{opt:F} and \eqref{opt:F_alt} are equivalent. Essentially, constraint \eqref{opt:F_alt_con} enforces that the uncertainty set must have one point at the boundary of the feasible set. By minimizing $\alpha$, formulation \eqref{opt:F_alt} seeks for the smallest uncertainty set $T(\alpha)$ that intersects with the boundary of the feasible set. This is the largest uncertainty set that can be enclosed in the feasible set.

Based on this property, a mixed-integer programming formulation is proposed that converts the equality constraint \eqref{opt:F_alt_con} to the Karush-Kuhn-Tucker (KKT) conditions of problem \eqref{opt:feasibility} \cite{grossmann1987active}. Doing so enforces that $\alpha$ must be large enough so that there exists one realization $\xi$ that lies on the boundary of the feasible set. The resulting formulation is:
\begin{subequations}
\label{opt:F_mip}
\begin{alignat}{2}
& F = \quad & \underset{\alpha,\xi,x,\lambda,s,y,\mu}{\min} \quad & \alpha \\ 
&& {\text{s.t.}} \quad & \sum_{j \in \mathcal{J}^I}\lambda_j = 1 \\
&&& \sum_{j \in \mathcal{J}^{E}} \mu_j \nabla_x h_j(x;\xi,\bar{\delta}) + \sum_{j \in \mathcal{J}^I} \lambda_j \nabla_x g_j(x;\xi,\bar{\delta}) = 0 \\
&&& h_j(x;\xi,\bar{\delta}) = 0, \quad j \in \mathcal{J}^{E} \\
&&& g_j(x;\bar{\delta}) + s_j = 0, \quad j \in \mathcal{J}^I \\
&&& \lambda_j \leq y_j, \quad j \in \mathcal{J}^I \\
&&& s_j \leq U(1-y_j), \quad j \in \mathcal{J}^I \\
&&& \lambda \geq 0, s \geq 0, \alpha \geq 0, y \in \{0,1\}^{|\mathcal{J}^I|} \\
&&& \xi \in T(\alpha) 
\end{alignat}
\end{subequations}
In problem \eqref{opt:F_mip}, $\mu$ and $\lambda$ are the Lagrange multipliers for the equality and inequality constraints, respectively. $s$ denote the slack variables for the inequality constraints. $y$ denote binary variables indicating whether inequality constraint $j$ is active or not. The constant $U$ is an appropriate upper bound for the slack variables $s$. When the constraints $g_j(\cdot)$, $h_j(\cdot)$ and the set $T(\alpha)$ are convex, the problem is a convex mixed-integer program (which can be solved efficiently). Any nonconvex constraint  $g_j(\cdot)$, $h_j(\cdot)$ or a nonconvex set $T(\alpha)$ will lead to a nonconvex mixed-integer program (which are more computationally intensive); moreover,  non-convex  formulations cannot guarantee feasible operation for every element of the uncertainty set (as in the convex case). 

\section{Case Study}
\label{sec:case_study}

In this section, we apply the flexibility analysis framework to demonstrate how it helps answer key questions relevant to quantifying flexibility provision in different settings. The flexibility analysis framework is implemented in \texttt{Julia}; code and data needed to reproduce the results are available at \url{https://github.com/zavalab/JuliaBox/tree/master/FlexQuantVL}. 

\subsection{Optimal Deployment of Spatial Flexibility}
\label{subsec:case1}

We demonstrate how to apply the flexibility analysis framework to answer the following question: What is the most efficient way to incorporate spatial flexibility in the system? In the context of virtual links, we reformulate the question as \textit{finding the optimal set of virtual links that maximizes the improvement in flexibility}. We consider a purely spatial case based on the modified IEEE-118 case with Active Power Increase (API) \cite{babaeinejadsarookolaee2019power} at a fixed time. The net demand at each node $D_n$ is treated as an uncertain parameter, and the nominal point is selected as the reference value $\bar{D}_n$ given in the case data. Thus, the hyperbox uncertainty set can be written as 
\begin{equation}
T(\alpha) = \{D \,|\, (1-0.5\alpha) \bar{D}_n \leq D_n \leq (1+0.5\alpha) \bar{D}_n,\, n\in\mathcal{N}\}
\end{equation}
Under this definition, a value of $\alpha = 1$ means that the system remains feasible for net demand values that are within 50\% deviation in both directions from the reference values. 

\begin{figure}[!htp]
    \centering
    \includegraphics[width=1\textwidth]{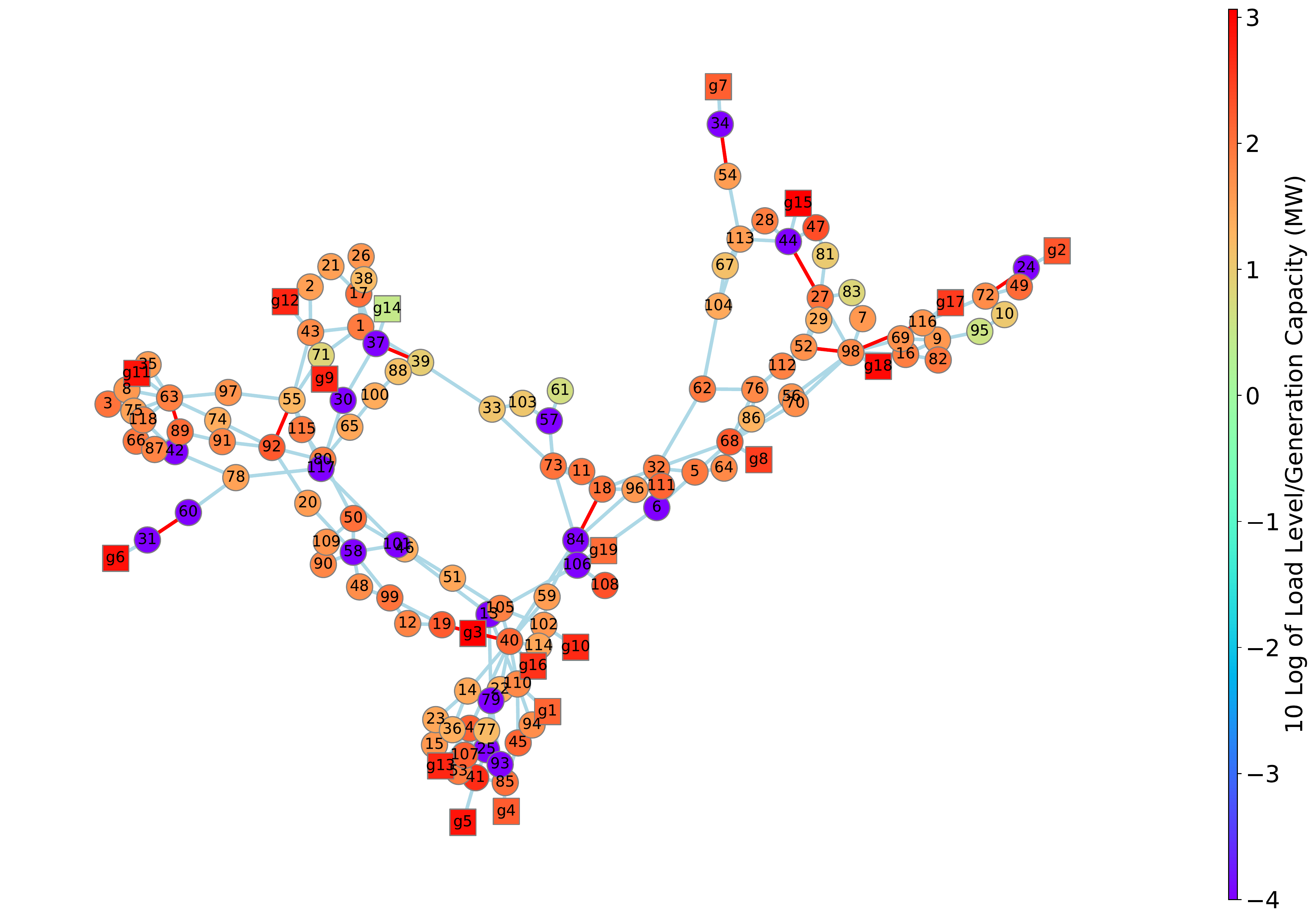}
    \caption{\small Network structure of the IEEE 118-bus case. Each circle denotes a bus, and each square denotes a generator attached to a bus. Transmission lines are shown between buses in lightblue. Red lines denote transmission lines with active capacity constraints in the base case. Colors denote the load level (for buses) or generation capacity (for generators) on a log scale.}
    \label{fig:case118_base}
\end{figure}

The IEEE 118-bus case is visualized in Figure \ref{fig:case118_base}; each bus (node) is represented by a circle, and each generator is represented by a square. The transmission network structure is shown by the lines. The color of each bus represents the amount of net demand in log scale; a positive value indicates that load is higher than generation at that location. A deep blue color means there is no load attached (pure generation is present at such node). A total of 99 out of 118 buses are connected to a load, while a few nodes are generation-only. The color of each generator represents the maximum generator capacity, also in log scale. We also note that some transmission lines are marked red, meaning that their capacity constraints are active for the base case (the lines are congested).  

\begin{figure}[!htp]
    \centering
    \includegraphics[width=0.6\textwidth]{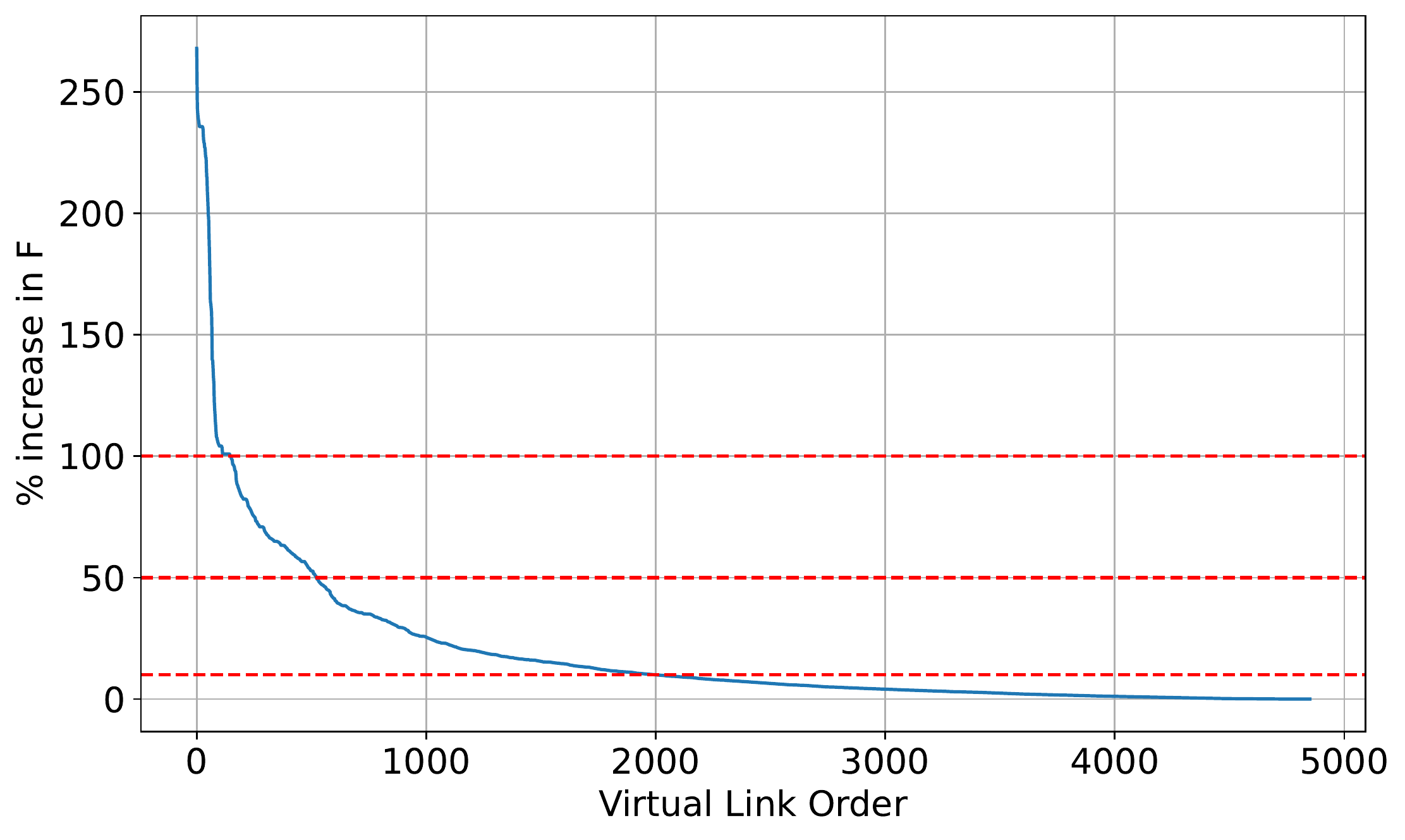}
    \caption{\small Percentage increase in flexibility from individual virtual links (in sorted order). Dashed horizontal lines denote the level of 10\%, 50\% and 100\% increase in flexibility (corresponding to 41.58\%, 10.76\%, and 3.09\% of all virtual links).}
    \label{fig:vl_order}
\end{figure}

We begin by exploring the benefit of adding a pair of virtual links (in both directions) between a pair of nodes to the base system. This assesses the benefit of converting a pair of loads  from being fixed to being spatially flexible. Each virtual link is added with a capacity of 30\% of the sending load (only 30\% of the load is shiftable). By doing this, we are effectively converting 30\% of the loads from being fixed to being shiftable to some other bus. This simulates the change of market mechanisms that account for flexible loads that are already in-place, but were unable to offer flexibility to the power grid/markets. With 99 different loads, there are $99 * 98 / 2 = 4851$ possible ways to add virtual links. We solve problem \eqref{opt:F_mip} for the flexibility index $F$ for all these alternatives.  Figure \ref{fig:vl_order} shows the percentage increase in $F$ compared to the base case for all 4851 possible pairs of virtual links, in decreasing order. The dashed lines correspond to the 100\%, 50\% and 10\% levels, which correspond to 3.09\%, 10.76\% and 41.58\% of all cases. It is clear that adding even a small amount of shifting can give a great boost in flexibility over the whole system, with a maximum increase of 267.97\% as measured by the flexibility index. However, the effect of adding such flexibility is highly sensitive to the location to which it is added, as in more than half of the cases the improvement will not be more than 10\%. Achieving greater benefits from flexibility thus requires careful selection for placement of flexible loads. 

\begin{figure}[!htp]
    \centering
    \includegraphics[width=0.9\textwidth]{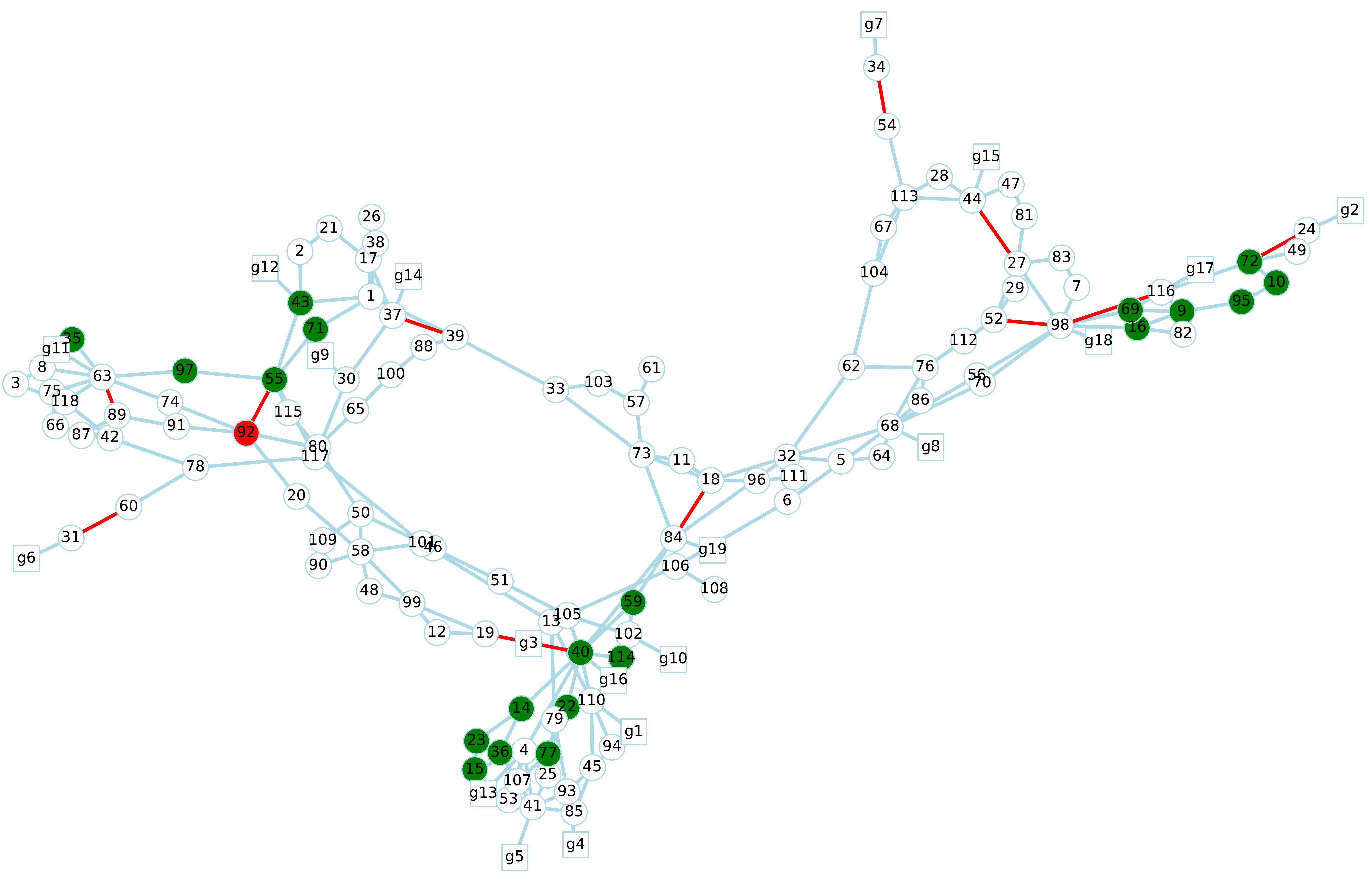}
    \caption{\small Nodes associated with the top 20 virtual links (shown in green). The red node denotes node 92, common node to all top 20 virtual links. Red lines mark transmission lines with active capacity constraints in the base case.}
    \label{fig:top_vl_nodes}
\end{figure}

\begin{figure}[!htp]
    \centering
    \includegraphics[width=1.05\textwidth]{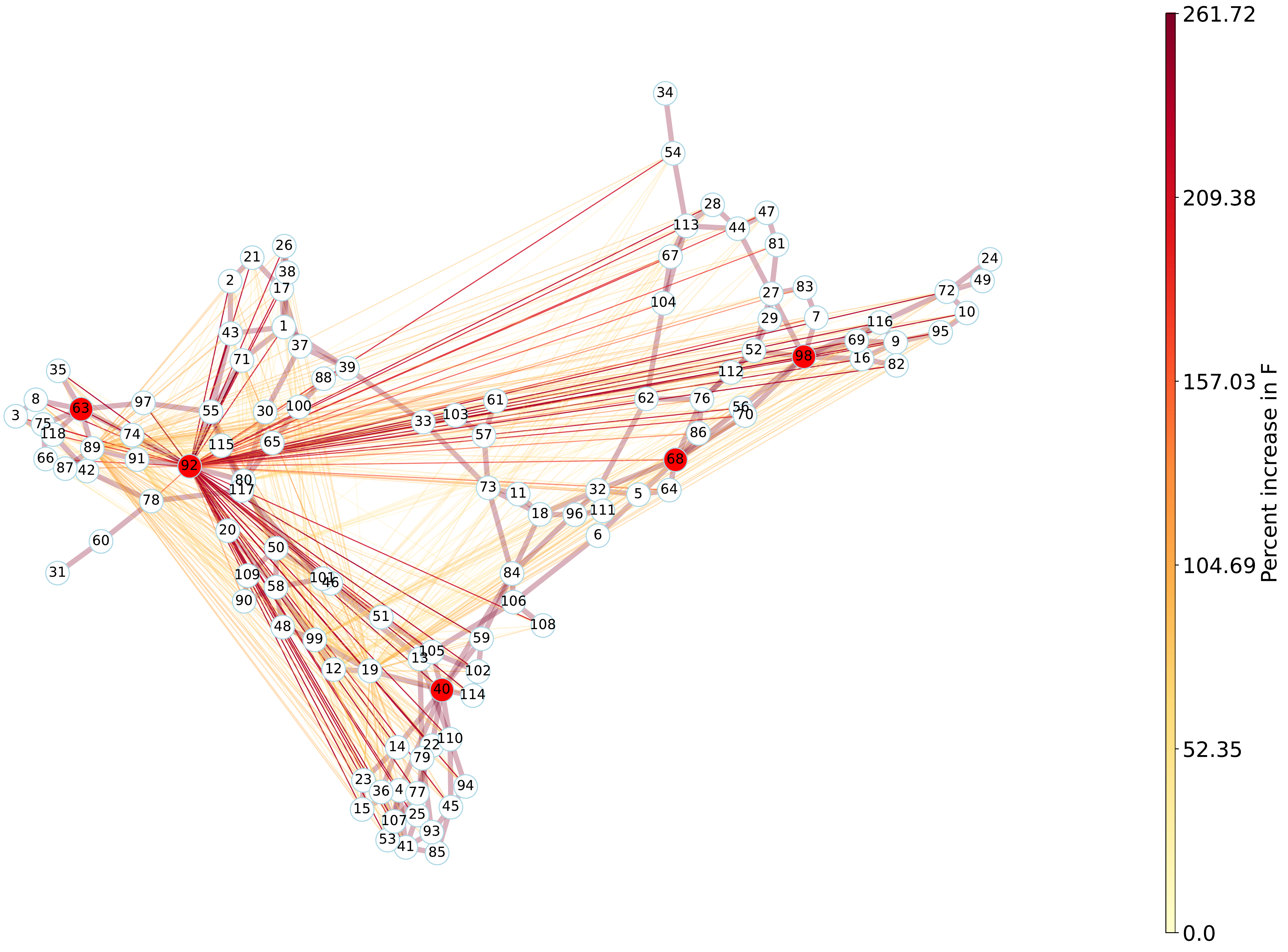}
    \caption{\small Visualization of flexibility index results for all virtual links (shown in thin solid lines). Darker lines denote higher flexibility index. Only virtual links with more than 20\% increase in flexibility index relative to the base case are shown. Five buses with the highest connectivity are marked red. Transmission network topology is shown by the transparent thick solid lines. }
    \label{fig:vl_all_F}
\end{figure}

An immediate question that follows is: How to choose the best few buses to deploy flexible loads? To address this question, we inspect the 20 virtual links with top flexibility increases in table \ref{table:vl_order}. Interestingly, all 20 virtual links are connected with bus 92, which is connected by an active transmission line and where the load level is relatively high. The topology of relevant buses are shown in figure \ref{fig:top_vl_nodes}. We observe that both short-range and long-range virtual links provide great contribution to flexibility. Specifically, short-range virtual links alleviate local congestion (as shown by the congested lines 92-55 and 63-89) by adjusting local load distribution in space. On the other hand, long-range virtual links adjust shift loads away from congested areas. We also observe a clustering behavior from nodes that contribute the most flexibility to the system. The clustering behavior becomes more pronounced in the virtual link visualization shown in Figure \ref{fig:vl_all_F}, where darker line color means higher flexibility index. The results indicate that adding long-range virtual links between these three set of nodes tends to contribute the most to increasing system flexibility.

\begin{table}[h!]
\centering
\begin{tabular}{| c c | c c |}
\hline
Bus 1 & Bus 2 & $\alpha$ & \% Increase\\
\hline
92 & 55 & 0.120 & 267.97 \\
92 & 71 & 0.118 & 261.72 \\
92 & 97 & 0.113 & 245.97 \\
92 & 40 & 0.112 & 243.40 \\
92 & 114 & 0.112 & 241.76 \\
92 & 22 & 0.111 & 241.46 \\
92 & 59 & 0.111 & 239.81 \\
92 & 14 & 0.111 & 239.10 \\
92 & 23 & 0.110 & 238.51 \\
92 & 77 & 0.110 & 238.02 \\
92 & 36 & 0.110 & 237.58 \\
92 & 15 & 0.110 & 236.94 \\
92 & 9 & 0.110 & 235.78 \\
92 & 72 & 0.110 & 235.78 \\
92 & 43 & 0.110 & 235.78 \\
\hline
\end{tabular}
\caption{Virtual links with top increase in flexibility.}
\label{table:vl_order}
\end{table}

\subsection{Economic Benefits of Flexibility Provision}

So far we have explored how virtual links contribute to the expansion of the operational feasible set of the market formulation. One key aspect that the previous analysis does not cover is the economic benefits of incorporating flexibility. This can be critical as, in most cases, incorporating flexibility might be expensive. In this subsection we address this issue by modifying the market formulation, so that economic quality of added flexibility is guaranteed to some level. This can be done by incorporating economic constraints to the market formulation. Here, we consider the following types of economic constraints:
\begin{itemize}
\item Total cost bound:
$$\sum_{t \in \mathcal{T}} \sum_{i \in \mathcal{S}} \alpha_{i,t}^p p_{i,t} + \sum_{v\in\mathcal{V}} \alpha_v^\delta \delta_v \leq (1+\epsilon) \psi_0 $$
\item Per-unit cost bound:
$$\sum_{t \in \mathcal{T}} \sum_{i \in \mathcal{S}} \alpha_{i,t}^p p_{i,t} + \sum_{v\in\mathcal{V}} \alpha_v^\delta \delta_v \leq (1+\epsilon) \frac{\psi_0}{\sum_{n,t}D_{n,t}}\sum_{n,t}d_{n,t} $$
\end{itemize}
where $\epsilon > 0$ is the additional fraction of cost that the market solution is allowed to admit, and $\psi_0$ is the optimal cost of the base case. The total cost bound constraints enforce that with the virtual links, the new solutions can only admit a total cost that is $\epsilon$ times higher than the optimal cost of the base case. On the other hand, the per-unit cost bound constraints enforce a similar bound, but on the average cost per unit of load served. 

\begin{figure}[!htp]
    \centering
    \includegraphics[width=0.8\textwidth]{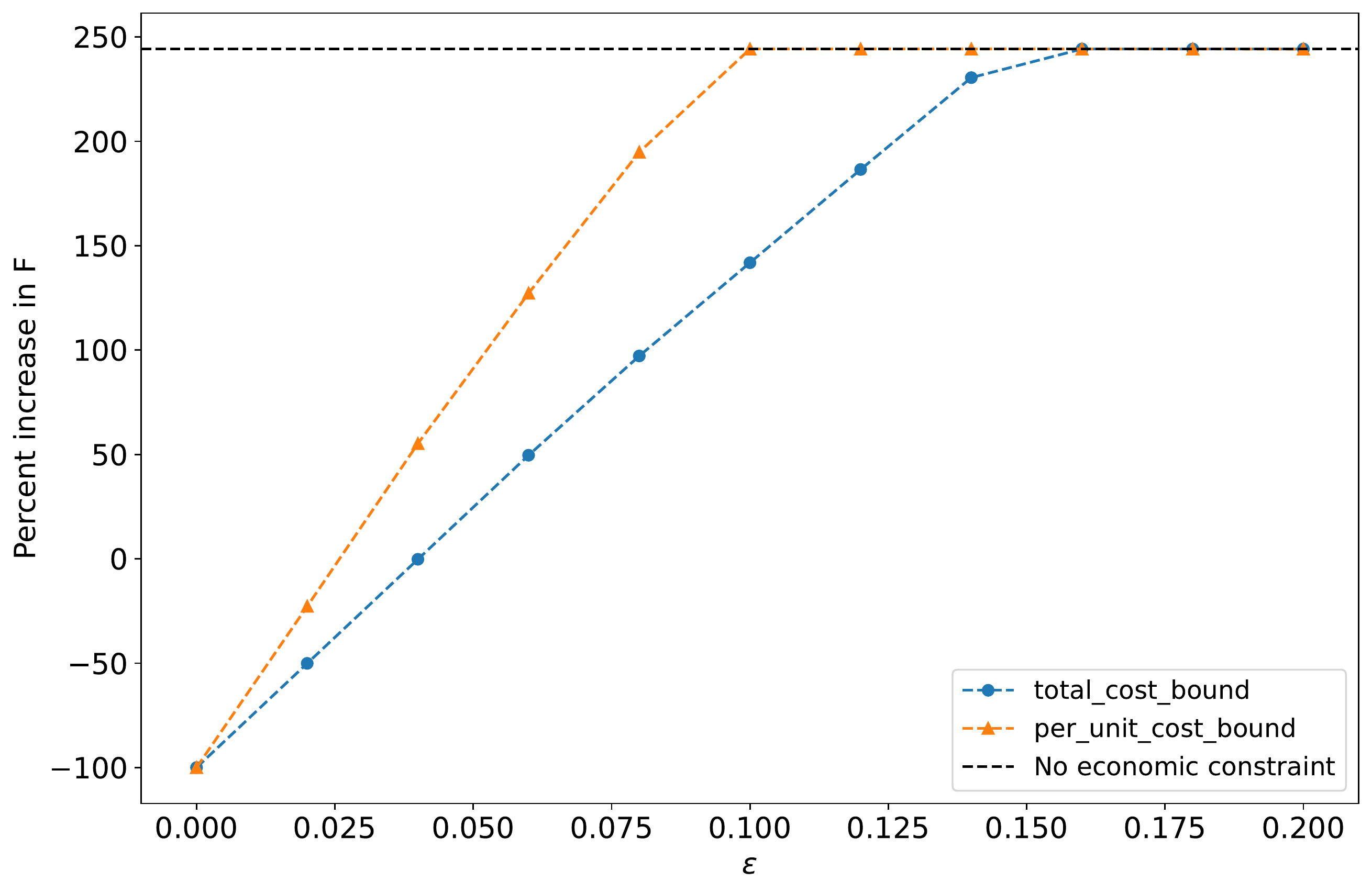}
    \caption{\small Percent increase in flexibility with respect to $\epsilon$, fraction increase in cost allowed. }
    \label{fig:econ_constraints_results}
\end{figure}

To explore the effect of adding economic constraints, we run the flexibility analysis for the complete network of virtual links within nodes $[92,9,40,45]$ with total cost bound in one case, and per-unit cost bound in the other case. For each case, we run the flexibility analysis multiple times with varying level of $\epsilon$.  Figure \ref{fig:econ_constraints_results} shows the experimental results with varying level of $\epsilon$. When $\epsilon = 0$, we obtain a relative increase of -100, meaning that the flexibility index is 0. This is expected, as we are enforcing that the cost must be exactly the same as the optimal cost of the base case and, therefore, only the base load level can be feasible. For both cases, the flexibility index increases with increasing $\epsilon$, as the economic constraints become more relaxed. Eventually when $\epsilon > 0.15$, we recover the percent increase in flexibility index with no economic constraints at all. This means that with a 15\% increase/relaxation in cost budget, the full benefit of added flexibility can be realized. Comparing the two trajectories, we notice the flexibility index values of total cost bound cases are lower bounded by those of per-unit cost bound cases for all levels of $\epsilon$. This is expected because the per-unit cost bound is supposed to be more relaxed compared to the total cost bound. The intuition behind this is that total cost bound is limiting the increase in cost incurred by having to satisfy more loads, whereas the per-unit cost bound only needs to bound the increase in average cost. 

\subsection{Optimal Flexibility Deployment of Temporal Flexibility}

\begin{figure}[!htp]
    \centering
    \begin{subfigure}[b]{0.45\textwidth}
    	\centering
	    \includegraphics[width=\textwidth]{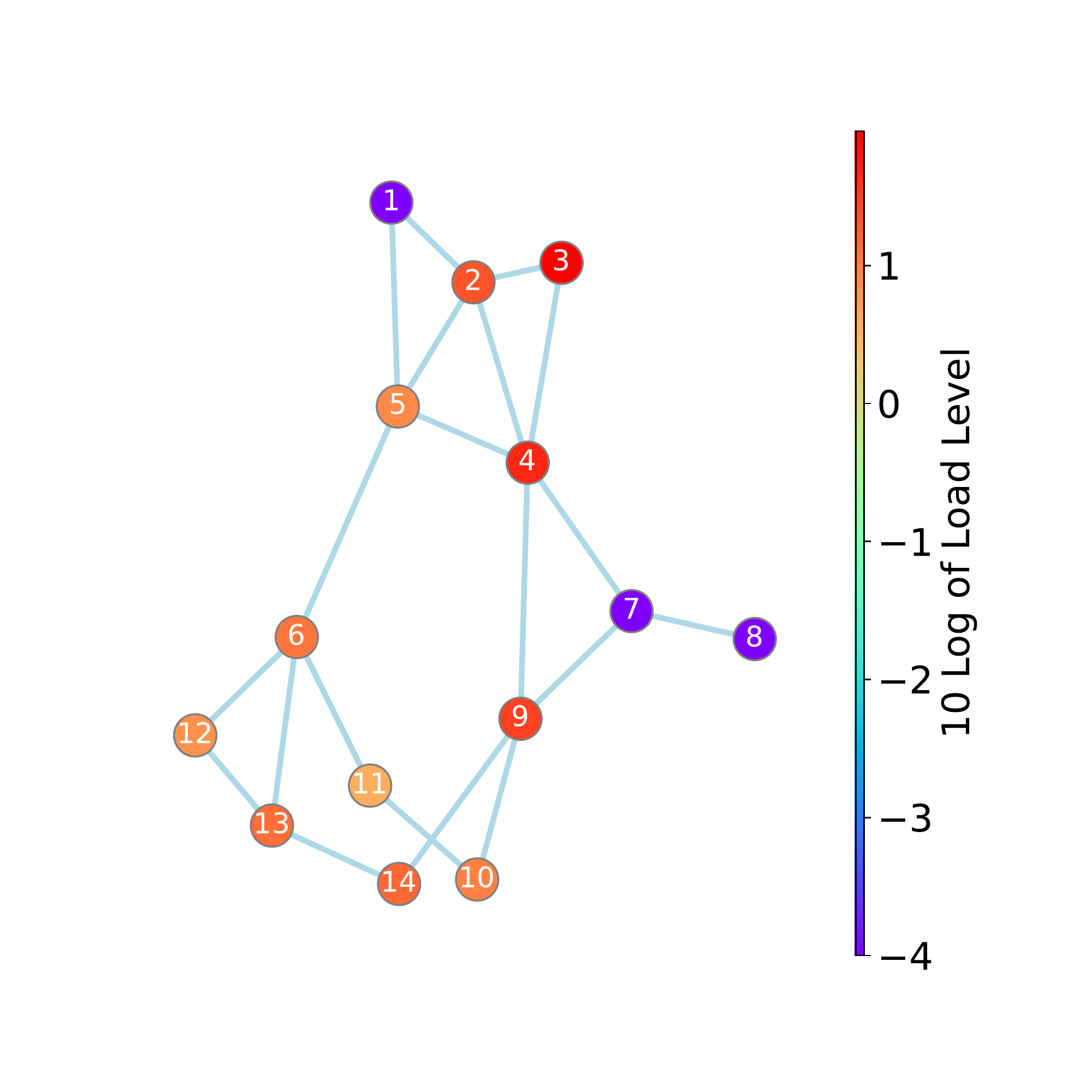}
    	\caption{}
    \end{subfigure}
    \hfill
    \begin{subfigure}[b]{0.45\textwidth}
    	\centering
	    \includegraphics[width=\textwidth]{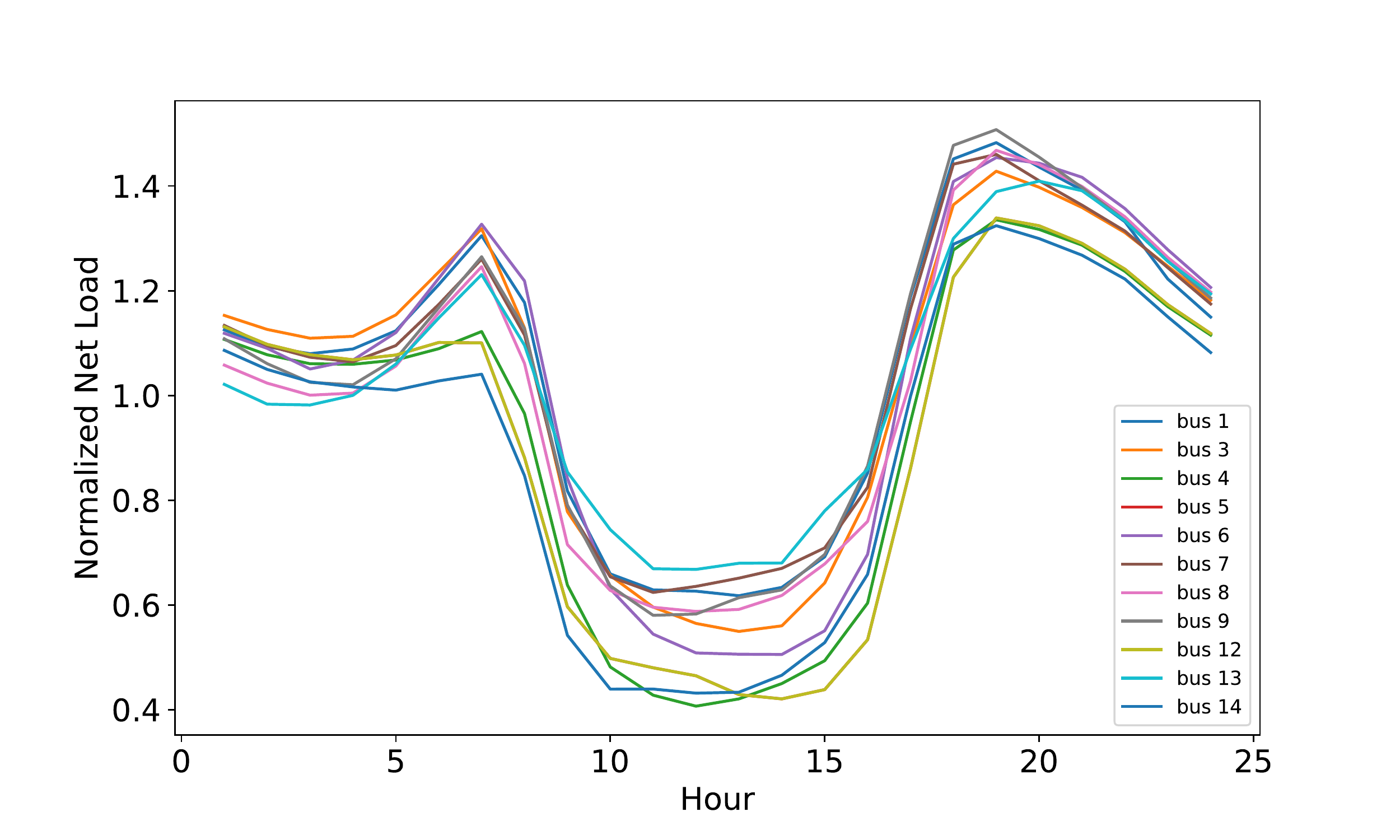}
    	\caption{\label{fig:duck_curve}}
    \end{subfigure}
    \caption{\small (a) 14-bus case network topology and (b) 24-hour demand profile. The system has a large generator installed at node 1 and a small generator installed at node 2.}
    \label{fig:14_bus_case}
\end{figure}

In this section, we demonstrate how to apply the flexibility analysis framework to analyze multi-period electricity markets with high renewable penetration. We consider a day-ahead market with time horizon of 24 hours and time resolution of 1 hr. The underlying power network is based on the 14-bus case study from the PGLib-OPF library \cite{babaeinejadsarookolaee2019power}. Each bus is installed with renewable energy, resulting in a duck-curve demand profile. Figure \ref{fig:14_bus_case} shows the network topology for the 14-bus case with base load level, and the normalized net load (relative to base load) profile for each node. The net load profiles are randomly generated based on CAISO net load data. By doing so, we assume that each node is connected to renewables, which give rise to the inverted duck-curve net load shape shown in figure \ref{fig:14_bus_case}\subref{fig:duck_curve}. We observe steep ramping of net load at round hour 8 and hour 16, which leads to need for flexibility to alleviate constraint due to generator ramping. 

For this system, we consider the question of choosing to install a storage system at one of the 14 buses, and the question of which pair of time points to add a pair of virtual links in both direction. For instance, by adding a virtual link to node $n$ between times $t_1$ and $t_2$, we are assessing the case where a storage system is installed at node $n$, and the storage offers power shifting services between time intervals $t_1$ and $t_2$. This means the storage offers to charge at $t_1$ and discharge at $t_2$, or charge at $t_2$ and discharge at $t_1$, and the market is free to clear in either direction.

\begin{figure}[!htp]
    \centering
    \includegraphics[width=0.8\textwidth]{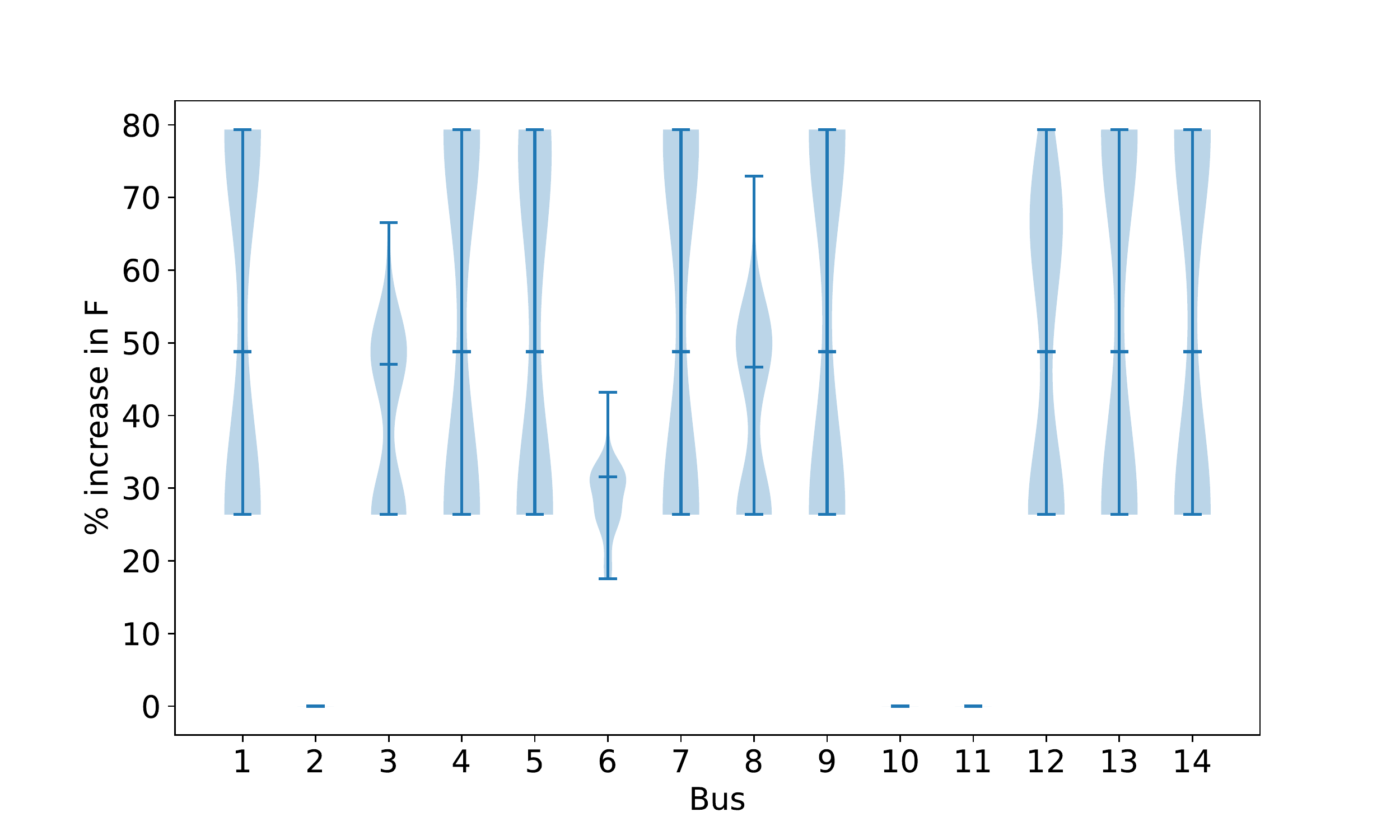}
    \caption{\small Percentage increase in flexibility index at different buses. Bars show the minimum, median, and maximum values for each node. Only data points with a positive percentage increase in flexibility are plotted.}
    \label{fig:violinplot}
\end{figure}
 
\begin{figure}[!htp]
\centering
\begin{subfigure}[b]{0.85\textwidth}
   \includegraphics[width=1\linewidth,trim={0 70 0 0},clip]{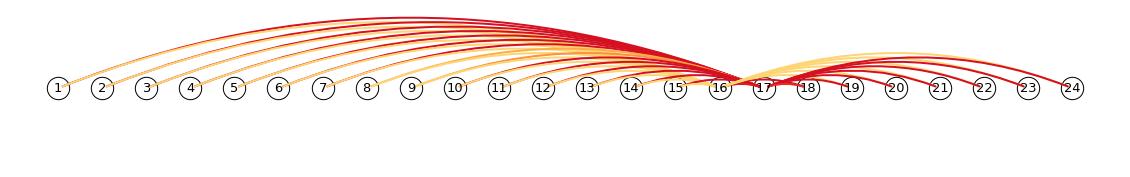}
   \caption{Node 1}
   \label{fig:temp_viz_1} 
\end{subfigure}

\begin{subfigure}[b]{0.85\textwidth}
   \includegraphics[width=1\linewidth,trim={0 70 0 0},clip]{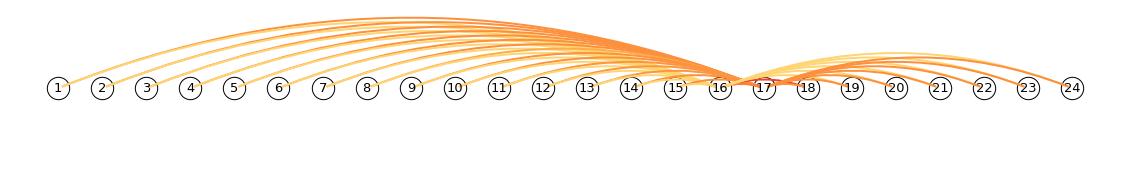}
   \caption{Node 3}
   \label{fig:temp_viz_3}
\end{subfigure}

\begin{subfigure}[b]{0.85\textwidth}
   \includegraphics[width=1\linewidth,trim={0 70 0 0},clip]{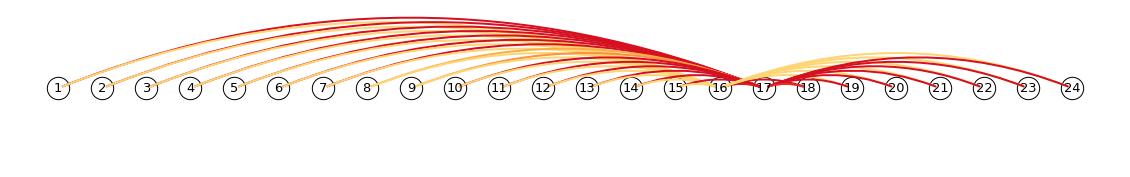}
   \caption{Node 4}
   \label{fig:temp_viz_4} 
\end{subfigure}


\begin{subfigure}[b]{0.85\textwidth}
   \includegraphics[width=1\linewidth,trim={0 70 0 0},clip]{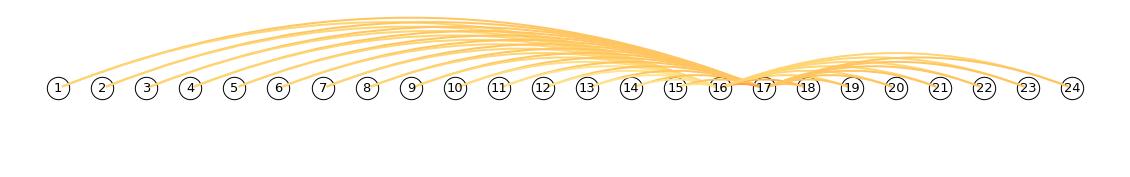}
   \caption{Node 6}
   \label{fig:temp_viz_6} 
\end{subfigure}




\begin{subfigure}[b]{0.85\textwidth}
   \includegraphics[width=1\linewidth,trim={0 70 0 0},clip]{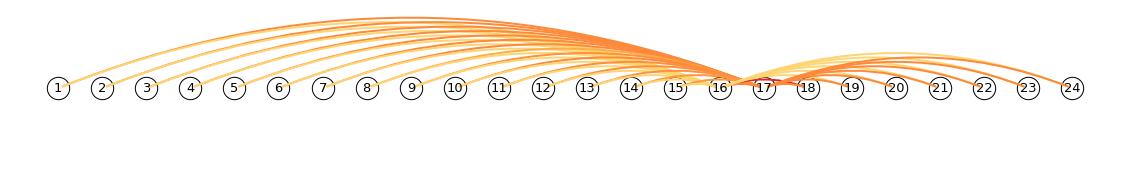}
   \caption{Node 8}
   \label{fig:temp_viz_8}
\end{subfigure}

\begin{subfigure}[b]{0.85\textwidth}
   \includegraphics[width=1\linewidth,trim={0 70 0 0},clip]{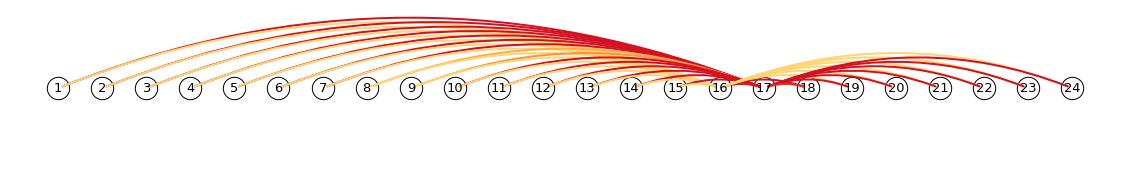}
   \caption{Node 13}
   \label{fig:temp_viz_13}
\end{subfigure}

\begin{subfigure}[b]{0.85\textwidth}
   \includegraphics[width=1\linewidth,trim={0 70 0 0},clip]{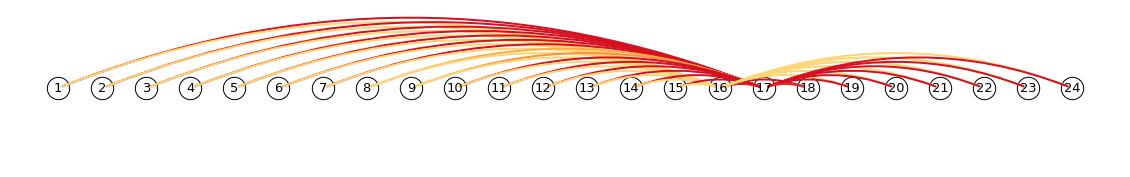}
   \caption{Node 14}
   \label{fig:temp_viz_14}
\end{subfigure}

\begin{subfigure}[b]{0.55\textwidth}
   \includegraphics[width=1\linewidth,trim={0 200 0 200},clip]{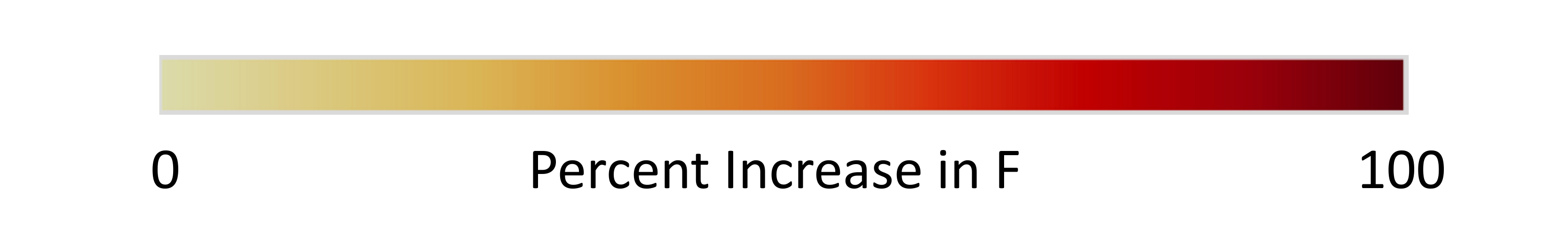}
   \label{fig:colorbar}
\end{subfigure}

\caption{Increase in flexibility index for selected nodes. Each circle represents a time interval. }
\label{fig:F_inc_by_nodes}
\end{figure}

Figure \ref{fig:violinplot} shows the increase of flexibility index for each node. We observe that adding one pair of virtual links to 11 of the 14 nodes results in the greatest increase in flexibility (80\%). Selecting the geographical nodes to add flexibility is still important, as we notice that adding virtual links to the wrong node (i.e. one of nodes 2, 10, and 11) results in no improvement of flexibility at all. However, this effect is not as obvious as the 118-bus case in section \ref{subsec:case1}. This is possibly a result of the fact that this case is small in terms of spatial size, meaning we have a small number of nodes and transmission lines, and there is not much congestion in transmission.

Figure \ref{fig:F_inc_by_nodes} shows the increase of flexibility index by each pair of virtual links for selected nodes. Only virtual links that achieve higher than 10\% increase in flexibility are visualized. Other nodes are not included either because no virtual links achieve higher than 10\% increase, or because they exhibit similar results to what are shown in the figure. We observe virtual links connecting hour 17 always achieve the highest increase in flexibility index for every node. This shows that the system has the largest temporal congestion around hour 17, which is expected as the net load profiles in figure \ref{fig:duck_curve} are steepest between hours 16 and 18.

\section{Conclusions and Future Work}
\label{sec:conclusion}
In this work, we propose a framework to quantify load shifting flexibility in electricity markets. We consider a clearing formulation with virtual links that capture space-time flexibility from shifting behavior. To quantify the improvement of flexibility from virtual links, our framework applies a systematic flexibility analysis based on the notion of flexibility index. This allows us to parameterize flexibility with a scalar quantity/index that can be computed efficiently using mixed-integer programming techniques. We perform several numerical case studies to demonstrate how this framework helps analyze benefits of flexibility investment to the electricity market. Our results show that even adding one pair of virtual links can bring about huge increase in flexibility index. This is true even with the consideration of economic viability of harnessing the added flexibility. We also show that the benefit of flexibility is highly dependent on the choice of location and time to add the flexibility. This is consistent with the general observation that power system operations are often constrained by transmission line capacity and generator ramping limits. For future work, we consider developing solutions that are more scalable than off-the-shelf mixed-integer solvers, which will help solve larger (and thus more realistic) cases. We also consider applying different types of uncertainty sets other than the hyperbox set considered in this paper. 


\section*{Acknowledgments}
We acknowledge support from the U.S. National Science Foundation under award 1832208.


\bibliography{refs}



\end{document}